\documentclass[aps,prc,preprint,superscriptaddress]{revtex4-1}

\usepackage{amsmath}
\usepackage{amssymb}
\usepackage{graphicx}
\usepackage{multirow}
\usepackage{hyperref}
\usepackage{wasysym}

\begin{document}

\title{Measuring the isotopic composition of trace elements in zircon: application to radiogenic molybdenum}%

\author{Adam J. Mayer}
\affiliation{Department of Physics and Astronomy, University of Calgary, Calgary, Alberta, Canada}
\email{ajmayer@ucalgary.ca}
\author{M. Wieser}
\affiliation{Department of Physics and Astronomy, University of Calgary, Calgary, Alberta, Canada}
\author{W. Matthews}
\affiliation{Department of Geoscience, University of Calgary, Calgary, Alberta, Canada}
\author{R.I. Thompson}
\affiliation{Department of Physics and Astronomy, University of Calgary, Calgary, Alberta, Canada}
\affiliation{TRIUMF, Vancouver, British Columbia, V6T 2A3, Canada}

\date{\today}

\begin{abstract}
Techniques have been developed to measure the isotopic composition of trace elements from matrices predominantly consisting of interfering isotopes. These techniques have been applied to measuring mass-independent fractionation of molybdenum in ancient zircon samples due to nuclear decays. Improvements to the digestion of large zircon samples and the ion exchange chemistry required to separate trace molybdenum from zirconium silicate are presented. The efficacy of TEVA ion exchange resin as a replacement for anion resin to improve separation efficiency and chemistry blank was studied. An algorithm was established to improve background, interference and mass bias corrections for the data analysis of Mo by multi-collector inductively coupled plasma mass spectrometry (MC-ICPMS). This enabled the isotopic composition measurement of $<$50 ng of Mo recovered from 500 mg ZrSiO$_4$ by reducing the Zr content by $>$9 orders of magnitude while retaining 63\% of the Mo. With 5 ng of Zr still remaining, novel data analysis techniques enabled $<0.1\permil$ precision on each Mo isotope. The relevance of this technique to extracting other atomic species is also discussed.
\end{abstract}

%\pacs{}
%\keywords{}

\maketitle
\newcommand{\regTM}{\textsuperscript{\tiny{\textregistered}}}  %Registered Trademark Symbol
\newcommand{\TM}{\textsuperscript{\tiny{TM}}} %Trademark symbol

\section{Introduction}

The measurement of isotopic compositions and delta values generally requires the separation of the element of interest from a complex matrix. For measurements by inductively coupled plasma mass spectrometry, the presence of a matrix can affect the measurement by changing the instrumental mass bias or suppressing the element of interest. Separation from the matrix is especially important when the matrix contains elements with interfering isotopes, in order to minimize corrections for isobaric interferences. 

The mineral zircon, ZrSiO$_4$, has been a staple of geochemistry for decades giving insights into the history of our planet\cite{Davis2003}. Zircon can remain a closed system over its lifetime, containing a chemical and isotopic fingerprint from the age when it was formed. In the early 1900s, scientists studying mass spectrometry recognized that zircon would be one of the best tools to use as a geochronometer. The discovery of the radioactivity of uranium and its eventual decay to lead would provide a tool to measure sample ages on a geological timescale, from thousands to billions of years. Zircon is especially suited to this measurement not only because of its resistance to chemical alteration, but also due to its high U concentration and the exclusion of lead from its structure during crystallization. Uranium-lead dating techniques\cite{Schaltegger2015} exploit this to measure the age of zircon minerals that are nearly as old as the Earth itself providing critical constraints on early Earth evolution processes.\cite{Harrison2009}. 

Not only is zircon valuable as a geochronometer, but it also contains traces of the environment in which it was formed. Zircon provides the only way to understand the environment on Earth from $>$4 Ga, since no whole rocks have survived that long\cite{Whitehouse2017}. By comparing ancient zircon's elemental and isotopic compositions to more recent zircons along with their host rocks, an understanding of the early Earth can be deduced. The analysis of rare earth elements\cite{Whitehouse2002} along with oxygen and hafnium isotopic compositions\cite{Whitehouse2017} have been used to infer details about early Earth evolution processes such as crustal formation and the presence of water. Recently, advances in analytical techniques have allowed for isotopic composition measurements of elements found at low concentrations such as lithium, which provides additional constraints on early Earth evolution\cite{Ushikubo2008}.

Beyond geological applications, the nature and age of zircons make them an ideal system to study very long-lived nuclear processes such as double-beta ($\beta\beta$) decay. For example, the isotope $^{96}$Zr, which has a natural abundance of 2.80\%\cite{DeLaeter2003}, has been the subject of ongoing studies\cite{Heiskanen2007} as a $\beta\beta$-decay candidate: $^{96}\text{Zr} \rightarrow $ $^{96}\text{Mo}+2e^{-}+2\bar{\nu}$. The high energy of this decay, $Q=3356.097(86)$ keV\cite{Alanssari2016}, makes it one of the best candidates for the observation of neutrinoless $\beta\beta$-decay, the detection of which would prove the neutrino to be its own anti-particle. Two previous studies of the $\beta\beta$-decay half-life yielded significantly different results: $0.94(32)\cdot10^{19}$ years\cite{Wieser2001} and $2.4(3)\cdot10^{19}$ years \cite{Argyriades2010}. The first was performed by a geochemical measurement of zircon, while the second was a direct counting measurement performed at the Neutrino Ettore Majorana Observatory (NEMO). To understand this discrepancy, a more careful geochemical measurement is of utmost importance.

A geochemical measurement of the $\beta\beta$-decay half-life is performed by measuring the amount of the decay product $^{96}$Mo as a mass-independent excess relative to the natural isotopic composition of Mo. The measured excess is used along with the mass ratio of Mo to Zr in the zircon sample to determine the relative amount of daughter product with the following equation:

\begin{equation}
\frac{N_d}{N_0}=\frac{m_\text{Mo}}{m_\text{Zr}}\frac{A_w(\text{Zr})}{A_w(\text{Mo})}\frac{C(^{96}\text{Mo})}{C(^{96}\text{Zr})}\delta(^{96}\text{Mo})
\label{eq:ndn0}
\end{equation}

where $N_d$ and $N_0$ are the numbers of daughter $^{96}$Mo and parent $^{96}$Zr atoms respectively, $m_\text{X}$ are the total masses of Mo and Zr in the sample, $A_w(\text{X})$ are the atomic weights of each element, $C(^{96}\text{X})$ are the natural isotopic abundances of the respective isotopes, and $\delta(^{96}\text{Mo})$ is the measured excess of $^{96}$Mo due to the daughter product. The half-life $T_{1/2}$ is then determined from this ratio and the average age, $t$, of the zircons with the following equation:

\begin{equation}
T_{1/2}=\frac{-t \ln{2}}{\ln{(1-N_d/N_0)}}
\label{eq:halflife}
\end{equation}

There are therefore three values which must be measured to determine the half-life: the $^{96}$Mo excess, the $m_\text{Mo}/m_\text{Zr}$ ratio (Mo:Zr), and the age of the zircons in the sample. The Mo isotopic composition can be measured by mass spectrometry but requires the sample to be digested, and the Mo must be separated from the Zr to eliminate isobaric interferences from $^{92,94,96}$Zr. This is particularly challenging as the Zr content is $>$10$^7$ times more than the Mo content. The large Mo:Zr ratio also makes it difficult to measure the Mo:Zr ratio itself due to the limited dynamic range of the ICPMS, measured to be $<$10$^6$:1 for Mo and Zr. The Zr and Mo content cannot be measured simultaneously, so the Zr content is measured prior to the first ion exchange, while the Mo content is measured after most Zr is removed. Lastly, the sample age is determined by U-Pb dating using LA-ICPMS.

In this study we performed a meticulous analysis and optimization of the measurement techniques required to measure Mo isotopic composition in zircon in order to perform a geochemical measurement of the half-life of the decay of $^{96}$Zr. Large quantities ($>$100 mg) of zircon were required to be digested in order to detect the decay product $^{96}$Mo (on the order of $\sim$10 pg radiogenic Mo and $\sim$10 ng common Mo per gram of zircon). The zircon digestion and ion exchange chemistry were optimized to ensure maximum Mo recovery and separation from Zr. Suppression of Zr was critical to the analysis, and a target of less than 10\% m(Zr)/m(Mo) remaining was required. The mass spectrometry and data analysis were analyzed and optimized to obtain the lowest possible measurement uncertainty from limited samples while enabling a quantification of the sources of uncertainty. In particular, the correction of background and isobaric interferences were improved.

\section{Samples, reagents and equipment}

Numerous zircon reference materials have been well-characterized by geochronologists and geochemists for use as calibration and validation reference materials\cite{Wiedenbeck1995}\cite{Black2004}\cite{Woodhead2005}. However, the limited availability of these reference materials and the large quantity of zircon required for testing and validation of the measurement method presented herein ($\sim$20 g used throughout experiments) precludes their use in this study. As such, separated zircon samples from the Yoganup Strand Line were obtained from a Westralian Sands Limited (WSL) mineral sand mining operations at Capel, Western Australia. Detrital samples often contain zircons spanning a wide range of ages and high-n detrital zircon geochronology by LA-ICP-MS was used to characterize the detrital zircon population in the sand. Dates for 510 individual zircon grains were obtained for an aliquot of zircon (WSL5655) employing methods similar to those of Daniels et al\cite{Daniels2017}. Individual zircon dates ranged from 150 Ma to nearly 3500 Ma and a mean age of 910(30) Ma was obtained for the sample.

Solutions were prepared with high purity reagents including Seastar\TM Baseline\regTM 47-51\% hydrofluoric acid (HF), Anachemia Environmental Grade Plus 32-35\% hydrochloric acid (HCl), and BDH Aristar\regTM Ultra 67-70\% nitric acid (HNO$_3$). Reagents were diluted with Milli-Q water purified to 18.2 M$\Omega\cdot$cm. Measurements were calibrated to dilutions of ICP standards: Specpure\regTM 1000 $\mu$g/g Zr and PlasmaCal\regTM 10000 $\mu$g/g Mo. All dilutions and other mass measurements were performed with a calibrated Mettler-Toledo AT201 analytical balance. 

Digestions were performed in a custom HF resistant Parr Instrument Company Model 4746 high pressure acid digestion vessel constructed from the high nickel Alloy 400 (Monel\regTM). Ion exchanges were performed with Eichrom\regTM TEVA resin (50-100 $\mu$m) and Eichrom\regTM analytical grade cation exchange resin (50Wx8, 100-200 mesh). Mass spectrometry measurements were carried out on a Thermo Scientific Neptune\TM multi-collector inductively coupled plasma mass spectrometer (MC-ICPMS) equipped with 9 Faraday cups, a secondary electron multiplier, and Multi Ion Counting (MIC) detectors. Samples were introduced through an Elemental Scientific Apex-Q desolvating nebulizer with a 130 $\mu$L/min PFA nebulizer.

\section{Digestion of zirconium silicate samples}

The digestion of small zircon samples is used frequently for geochemical analysis \cite{Potts2015}, especially in zircon dating by TIMS \cite{Mattinson2005}. Zircon is resistant to most acids, requiring high temperature HF for digestion. Concentrated HF is sometimes mixed with other acids such as nitric, sulfuric or boric acid to assist with digestion of the silicate matrix. High temperature digestion has been demonstrated to achieve close to 100\% digestion at 200-250 $^\circ$C for $>$24 hours.\cite{Mattinson2005,Wiedenbeck1995}

The digestion of the sample for this project was particularly challenging due to the large amount ($>$100 mg) of zircon being processed. Large sample sizes were required in order to have a detectable amount of the $^{96}$Mo decay product. Further, the HF acid digested sample had to be evaporated and fully re-dissolved in HCl for ion exchange separation. These requirements increased the likelihood that silicates would re-crystallize which could interfere with the recovery of trace elements.

Tests of recovery from the digestion were performed with 0.1-1.0 g aliquots of WSL5655 zircon, in 10 mL of hydrofluoric acid. Prior to digestion, zircon samples were purified by S.G. Frantz\regTM magnetic mineral separation (non-magnetic recovered at 1.8 A, 1$^\circ$), then washed in concentrated aqua regia overnight. This step minimized contamination from other minerals such as titanite (CaTiSiO$_5$) which were found to contain a large amount of Mo. Digestions were performed by loading the digestion vessel with the zircon and acid, then heating in an oven at 220 $^\circ$C for 3-4 days to achieve maximum dissolution. Zirconium recovery was tested by measuring the Zr signal intensity of a diluted (10000:1) sample against 100 ppb Specpure\regTM zirconium standard with the Neptune MC-ICPMS.

Digesting in 10 mL of 48\% HF yielded recoveries between 50-100\%, depending on the amount of zircon being processed. It was found that no more than 0.5 g of zircon could be successfully digested at once. Samples were then evaporated to dryness under a heat lamp then re-dissolved in 5 mL 2.75 M HCl for ion exchange. Solutions that contained visible crystallized silicate had lower Zr recovery during re-dissolution than those that achieved full digestion. Pipetting the dissolved solution from any crystallized silicate allowed for 95-100\% re-dissolution in HCl. The addition of nitric or boric acids was not found to improve the digestion or re-dissolution recovery. The addition of nitric acid was found to lower digestion recovery by diluting the HF acid.

Total Zr recovery of $>$75\% was achieved by digesting $\sim$0.5 g of zircon in 10 mL of 48\% HF acid at 215 $^\circ$C for 96 hours followed by evaporation of the silicate-free solution and re-dissolution in 2.75 M HCl, as shown in Table \ref{tab:Digest-Table}.

\begin{table}[t]
	\centering
		\begin{tabular}{llll}
		\hline
		  & 5655-0 & 5655-1 & 5655-2 \\
		\hline
		\textbf{Digestion} &  &  &  \\
		Mass Zircon (mg) & 383.0 & 573.8 & 489.5 \\
		Mass w/ HF (g) & 9.9756 & 12.6130 & 11.6684 \\
		$\left[\text{Zr}\right]$ Expected (mg/g) & 19.2 & 22.75 & 20.98 \\
		$\left[\text{Zr}\right]$ Measured (mg/g) & 20(1) &  - &  - \\
		\hline
		\textbf{Re-dissolution} &  &  &  \\
		Mass w/ HCl & 12.2786 & 12.5062 & 10.7136 \\
		$\left[\text{Zr}\right]$ Expected (mg/g) & 15.60 & 22.94 & 22.85 \\
		$\left[\text{Zr}\right]$ Measured (mg/g) & 15.3(8) & 23(1) & 10.3(5) \\
		\hline
		\end{tabular}
	\caption{Results of three digestions of WSL zircons. Sample 5655-0 demonstrated complete recovery during digestion, and samples 5655-0 and -1 demonstrated complete recovery during re-dissolution. Sample 5655-2 lost about half of the sample due to re-crystallization after digestion.}
	\label{tab:Digest-Table}
\end{table}

\section{Ion exchange separation of Mo from Zr}

\begin{figure}[htp]
	\centering
		\includegraphics[width=0.55\linewidth]{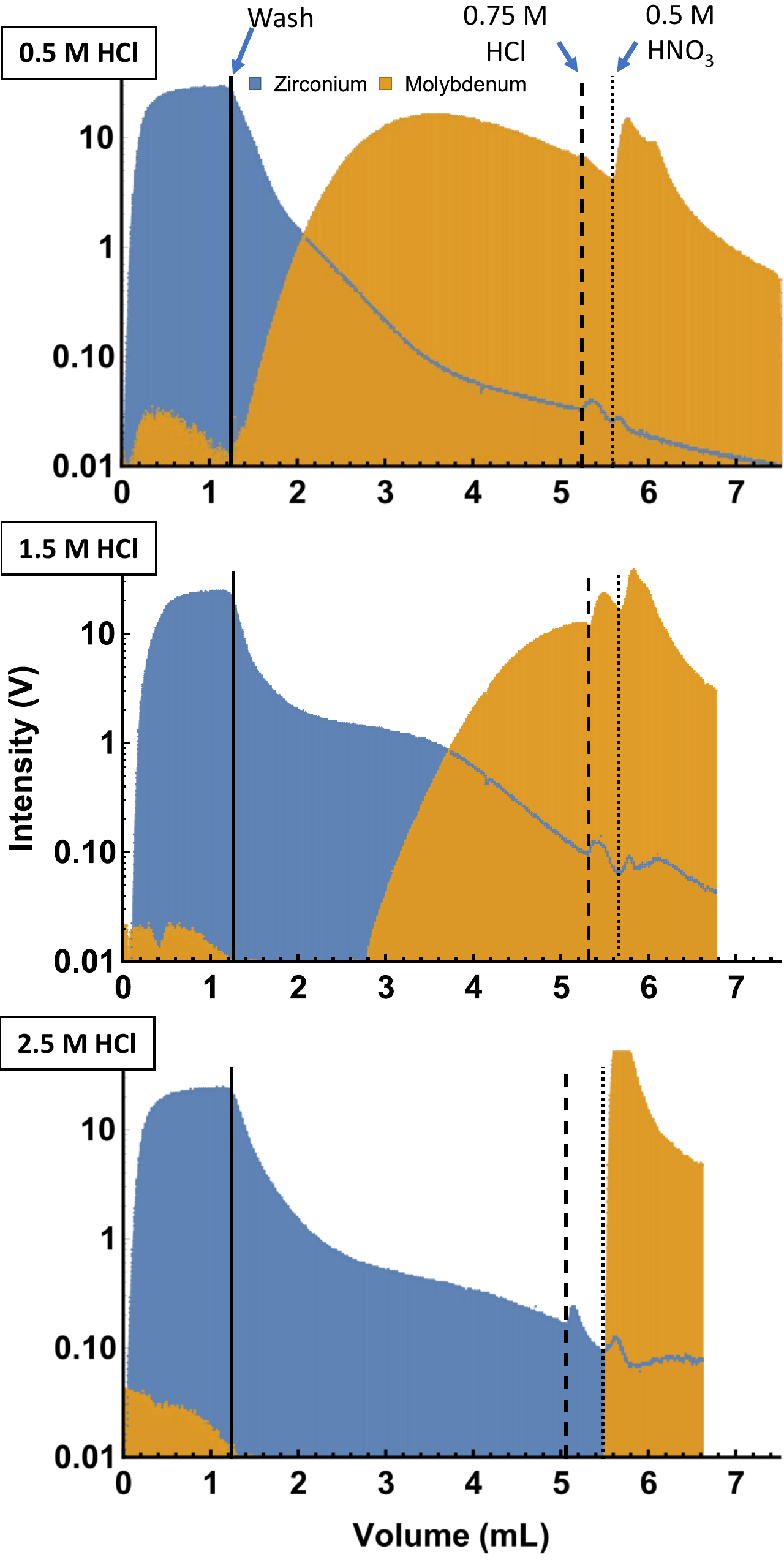}
	\caption{\footnotesize{Log-scale elution curves for solution passed through TEVA ion exchange resin at 130 $\mu$L/min into ICPMS. Solutions containing 1 ppm Zr and Mo in 0.5 M, 1.5 M, and 2.5 M HCl introduced at start, followed by wash in HCl at the same concentration, marked by a solid line.  At dashed line 0.75 M HCl is introduced, and at dotted line 0.5 M HNO$_3$ is introduced. Note, Mo signal prior to wash is due to tailing from the previous measurement.}}
	\label{fig:Elution-Curves}
\end{figure}

Purification of molybdenum from various matrices has been demonstrated to achieve $\sim$3-4 order of magnitude reduction in contaminants using anion and cation ion exchange resins \cite{Siebert2001}. However, a much more robust method was needed for this work as molybdenum must be separated from 7-8 orders of magnitude more zirconium to minimize the isobaric interferences on $^{92,94,96}$Mo. Further, the chemistry blank must be minimized to avoid increasing the $<$50 ng natural Mo content which would suppress the relative magnitude of the nuclear decay excess. Improved blanks were achieved by using TEVA resin (50-100 $\mu$m) for anion exchange as it was found to perform quite well with low column volumes and lower molarity acids. The resin was characterized by performing online elution measurements of a solution containing 1 ppm each of Mo and Zr in 0.25-3.0 M HCl. The solution was pumped at 130 $\mu$L/min through 0.5 mL of resin, then into the MC-ICPMS using an ESI Apex-Q desolvating nebulizer. The signal intensities for $^{91}$Zr and $^{98}$Mo were simultaneously measured and are shown in Figure \ref{fig:Elution-Curves}. Complete Mo retention was achieved at 2.0 M HCl while no evidence of increased Zr retention was observed up to 3.0 M HCl. It is important to note that Mo loaded with Zr is significantly delayed, even in 0.5 M HCl. This combined with the its broadened peak width demonstrated that the transport of Mo is affected by the resin. This effect is mitigated by using 0.5 M HNO$_3$ during Mo elution which significantly increases the elution rate as seen after the dotted lines in Figure \ref{fig:Elution-Curves}.

The optimal concentration of HCl for the ion exchange was found to be 2.75 M, resulting in high Mo retention and low Zr retention on the column while allowing for lowered acid concentration in case the acid is buffered during sample re-dissolution. Approximately 2 orders of magnitude reduction in Zr was achieved after 2000 s, corresponding to 5 mL of wash, with 3 orders achieved after 10 mL and diminishing returns after this point. Near complete ($>$95\%) recovery of Mo was achieved with 5 mL of 0.5 M HNO$_3$. To achieve sufficient separation, 2-3 complete passes through fresh resin was necessary.

In addition to the TEVA separation, a cation exchange resin was used to purify Mo from other elements, in particular iron which produces an interference on $^{96}$Mo due to $^{56}$Fe$^{40}$Ar$^+$. The interfering signal intensity was around 2.4 mV/ppm Fe, compared to 35 V/ppm Mo at $^{96}$Mo. This was enough to cause a significant non-radiogenic $^{96}$Mo excess in high Fe, low Mo samples. TEVA-exchanged samples were dried and re-dissolved in 0.5 M HCl. At this concentration, Fe and Zr are retained by the column while Mo is eluted. The solution was pumped through a 0.5 mL cation exchange column at 0.5 mL/min, and near-complete recovery ($>$95\%) was achieved with an additional 3-5 mL of 0.5 M HCl. This process was repeated several times to further minimize the Zr content. Finally, the purified Mo samples were dried and re-dissolved in 0.5 M HNO$_3$ for analysis by MC-ICPMS. Column blanks given as concentrations relative to the amount of solution passed through the column were measured to be less than 0.05 ng/g Zr and 0.01 ng/g Mo. Total analytical blank was measured to be $<$2 ng Mo, mostly coming from the acid digestion.

Four ion exchanges were performed on two separate digestions of the WSL zircons. Two ion exchanges using TEVA resin were performed to remove the bulk of the Zr from the sample as the extremely high concentration of Zr would quickly overload the cation columns leading to poor separation. Subsequently, two cation exchange separations were performed to capture the remaining Zr and other interfering elements such as Fe. As shown in Table \ref{tab:IonX-Table}, more than nine orders of magnitude reduction in Zr was achieved while retaining 63\% of the zircons' Mo. It was not possible to measure the initial Mo content prior to the ion exchange due to the very low concentration relative to Zr.

\begin{table*}[t]
	\centering
		\begin{tabular}{lllllll}
		\hline
		 & Initial & TEVA-1 & TEVA-2 & CAT-1 & CAT-2 & Proportion Remaining \\
		\hline
		5655-1 Mass-Zr & 287 mg & 180 $\mu$g & 145 ng & 19 ng & 0.3 ng & $8.7\cdot10^{-10}$ \\
		5655-1 Mass-Mo &  - & 56 ng & 45 ng & 42 ng & 35 ng & 63\% \\
		&  &  &  &  &  &  \\
		5655-2 Mass-Zr & 110 mg & 59 $\mu$g & 43 ng & 15 ng & 0.1 ng & $9.1\cdot10^{-10}$ \\
		5655-2 Mass-Mo &  - & 260 ng & 225 ng & 239 ng & 165 ng & 63\% \\
		\hline
		\end{tabular}
	\caption{Recoveries after four consecutive ion exchanges, two using TEVA resin and two using cation resin. Results demonstrate a 9 order of magnitude reduction in the Zr content while maintaining 63\% recovery of molybdenum. The large variation of Mo content was due to the accessory mineral titanite that was found to contain a much higher Mo concentration than the zircons. Uncertainties in mass measurement are 5\%.}
	\label{tab:IonX-Table}
\end{table*}

\section{Mo isotopic composition measured by MC-ICPMS}

The low quantity of Mo recovered from the zircons, often $<$50 ng, required careful optimization to achieve high precision due to multiple isobaric interferences. The mass-independent excess of $^{96}$Mo was used to determine the amount of decay product from the $\beta\beta$-decay of $^{96}$Zr $\rightarrow$ $^{96}$Mo. To measure this excess, mass-dependent fractionation, both natural and that which is introduced by the chemistry and instrument, must be corrected. Typically, this would be done by correcting to the accepted value for an interference free isotope amount ratio such as n$(^{97}$Mo)/n($^{95}$Mo). However, natural $^{238}$U contained in the zircons decays by spontaneous fission to $^{95,97,98,100}$Mo affecting the expected isotope amount ratio. This leaves n($^{94}$Mo)/n($^{92}$Mo) as the only Mo isotope amount ratio unaffected by mass-independent fractionation. It was therefore critical that any remaining Zr be accurately measured to correct for interferences from $^{92,94,96}$Zr. Ruthenium was also monitored due to interferences from $^{96,98,100}$Ru.

Samples were naturally aspirated into an ESI Apex-Q desolvating nebulizer at an uptake rate of $\sim$130 $\mu$L/min. $^{92-98}$Mo isotopes were measured in cups L2-H3 on $10^{11}$ $\Omega$ resistors, while $^{90,91}$Zr isotopes were measured in cups L4 and L3 with $10^{12}$ $\Omega$ resistors. $^{99}$Ru was monitored on an ion counter (IC6). $^{100}$Mo could not be measured simultaneously due to the relative cup positions of the interferences and was not included in the analysis. Each sample was measured with 60 cycles of 2 second integrations. The integration time was chosen to limit the effect of occasional spikes in the background intensity of interfering zirconium isotopes from the desolvating nebulizer.

The measurement sessions included measurements of the Mo ICP standard diluted to 10-200 ppb, the Zr ICP standard diluted to 50-100 ppb, and mixtures of the two standards were used to verify the performance of the interference correction. All samples were aspirated in 0.5 M HNO$_3$ and all measurements were preceded by blank measurements of 0.5 M HNO$_3$ to correct for other constant spectral interferences, such as $^{40}$Ar$_2$$^{16}$O$^+$. Every few measurements were bracketed by a measurement of the 200 ppb Mo working standard to provide an approximate correction for drift in instrumental mass bias. Once the performance of the laboratory standards had been verified, the purified Mo from zircons was measured with a concentration of $\sim$50 ppb in 1 mL of 0.5 M HNO$_3$.

\subsection{Data analysis}

A data analysis algorithm was developed using Mathematica\regTM to process the data. First, three filters were applied to the data: one to remove data with large spikes in Zr intensity, one to apply a 2$\sigma$ outlier test, and one to ensure the signal intensity was $>$75\% of the maximum. The latter was applied to maximize sample efficiency by starting data collection immediately when sample uptake is started, and when the sample runs out before data collection is finished. Next, a blank correction was applied by subtracting the average intensities of the HNO$_3$ blank measurements from the sample measurements, line by line.

The data were then corrected for Zr and Ru interferences using the average measured intensities at $^{90}$Zr and $^{99}$Ru for each sample. Ru content in the zircon Mo samples was $<$0.1\% relative to Mo. The Ru correction was applied using the IUPAC-published isotopic composition\cite{DeLaeter2003}, with an exponential fractionation correction applied to match the composition to the standard-sample bracketing determined fractionation, e.g.:

\begin{equation}
r\left(\frac{^{96}Ru}{^{99}Ru}\right)_{cor}=r\left(\frac{^{96}Ru}{^{99}Ru}\right)_{pub}\left(\frac{m_{96}}{m_{99}}\right)^\alpha
\end{equation}

where $r_{cor}$ and $r_{pub}$ are the fractionation corrected and IUPAC-published isotope abundance ratios, $m_x$ is the atomic mass of the $x$th isotope, and $\alpha$ is the fractionation exponent.

The effect of the Zr interference correction required more careful analysis as the Zr content is as much as 10\% relative to Mo. As most of the Zr was removed by ion exchange, the remaining Zr was significantly fractionated, which affected the correction. The measured $^{91}$Zr/$^{90}$Zr of each sample was used to apply an exponential fractionation factor to the Zr correction with a limit set at $\alpha = \pm 0.2$ for the fractionation exponent. This limit, corresponding to $^{91}\text{Zr}/^{90}\text{Zr}= \pm 2\permil$, was included to restrict the size of the fractionation correction on samples with only background Zr levels such as Mo ICP standards. Less than $\pm 1 \permil$ Zr fractionation was measured for samples containing Zr. The zirconium isotopic composition of the ICP standard was measured in the same session on the same cup configuration, which improved the accuracy of the Zr correction and allowed the $^{94}$Mo/$^{92}$Mo isotope ratio to be used to correct for mass-dependent fractionation.

\section{Measurement results}

\begin{figure}[htp]
	\centering
		\includegraphics[width=0.65\linewidth]{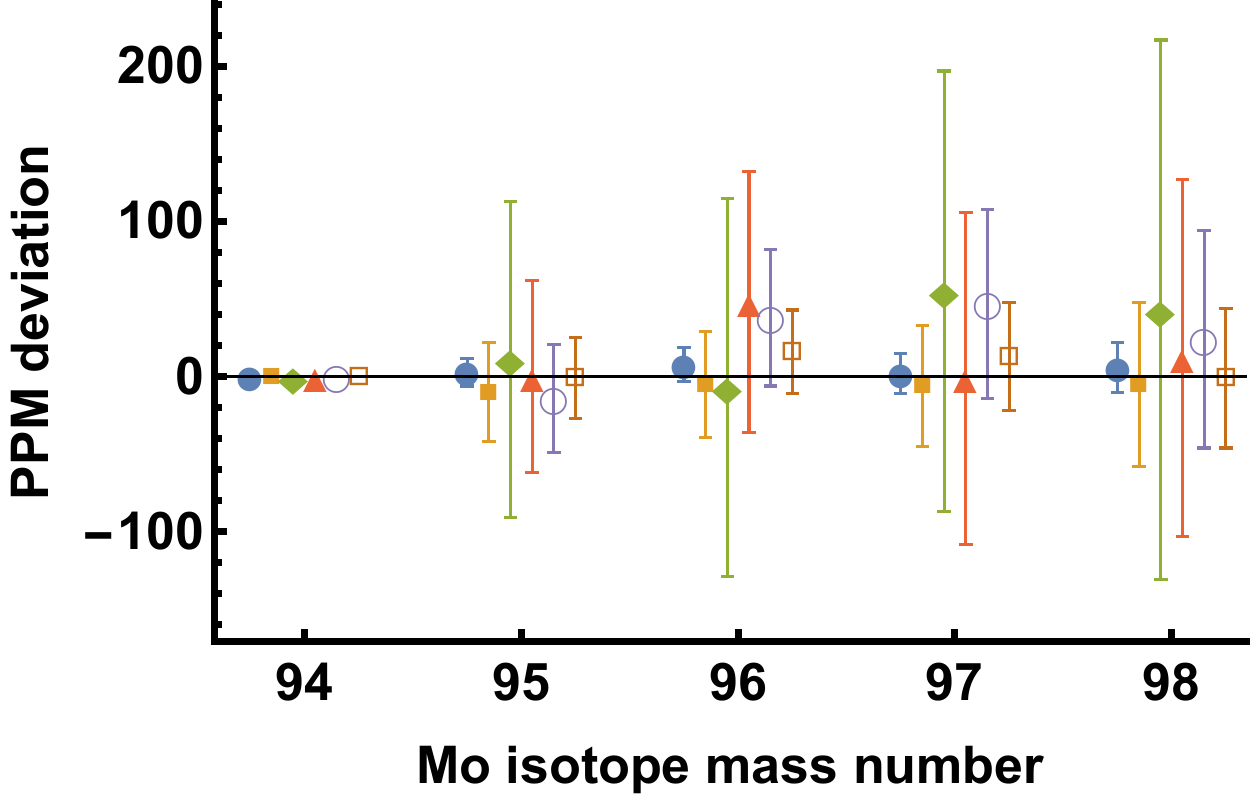}
	\caption{Measured isotopic compositions normalized to n($^{94}$Mo)/n($^{92}$Mo): Mo ICP standard (200 ppb - filled circle, 50 ppb - filled square, 10 ppb - filled diamond), mixture of 50 ppb Mo and 5 ppb Zr (filled triangle), and purified Mo from two WSL-5655 zircon samples (hollow circle and square). Error bars at 1$\sigma$ are included.}
	\label{fig:Delta-plot}
\end{figure}

To demonstrate the effectiveness of these techniques we look at typical results. A zircon sample from Westralian Sands Limited (WSL-5655) was measured along with a set of Mo and Zr standards and mixtures. Mo solutions with lower concentration and a synthetic mixture with higher Zr content than the Mo recovered from the zircons were used to verify accuracy of the data processing. As shown in Table \ref{tab:Delta-Table}, all measurements were within uncertainty of the initial lab standard measurement. Uncertainties ranged from tens to a hundreds of ppm, and depended predominantly on the success of the Zr correction. Since the mass bias correction is based on the $^{94}$Mo/$^{92}$Mo ratio, an increased uncertainty in the $^{92,94}$Zr correction leads to a magnified uncertainty in the higher mass difference isotope ratios. This is most evident in the lowest concentration Mo standard measurement. It is also evident from the fractionation trends seen in Figure \ref{fig:Delta-plot} that the largest deviations are due to uncertainty in the mass bias correction.

\begin{table*}[t]
	\centering
		\begin{tabular}{llllllll}
		\hline
		Sample ID & $^{90}$Zr (V) & $^{95}$Mo (V) & $\delta^{95}$Mo & $\delta^{96}$Mo & 	$\delta^{97}$Mo & $\delta^{98}$Mo & N \\
		\hline
		Mo 200 ppb & 0.0006 & 7.4 & 3(10) & 8(12) & 2(13) & 6(17) & 5 \\
		Mo 50 ppb & 0.0002 & 1.8 & -10(40) & -10(50) & -10(50) & -10(70) & 2 \\
		Mo 10 ppb & 0.0002 & 0.4 & 10(100) & -10(120) & 60(140) & 40(170) & 1 \\
		Mo 51 Zr 5 & 0.5516 & 1.9 & 0(50) & 50(60) & 0(80) & 10(100) & 2 \\
		WSL-5655 (1) & 0.0045 & 0.9 & -10(40) & 40(50) & 50(60) & 20(70) & 3 \\
		WSL-5655 (2) & 0.0011 & 3.2 & -1(23) & 16(26) & 13(29) & 0(40) & 3 \\
		\hline
		\end{tabular}
	\caption{Delta Values ($^{X}$Mo/$^{92}$Mo, ppm) relative to initial Mo standard reference measurement. Fractionation is corrected to n($^{94}$Mo)/n($^{92}$Mo) of initial Mo ICP standard measurement. Note that all delta-values, including the high-Zr synthetic mixture are within 1$\sigma$ uncertainty of 0 ppm, demonstrating the efficacy of the zirconium correction.}
	\label{tab:Delta-Table}
\end{table*}

The results of these relatively high Zr and low Mo samples demonstrate the accuracy of the data processing algorithm. The Zr interference correction is successfully applied even when the Zr concentration is 10\% of the Mo concentration. The uncertainty is well characterized, with no results lying outside the expected uncertainty range of zero. This establishes confidence in the results of the two independently processed zircon samples of WSL-5655, which demonstrated no resolvable evidence of nuclear processes leading to mass independent fractionation on any Mo isotopes.

\section{Limit on the half-life of $^{96}$Zr}

The half-life of the $\beta\beta$-decay of $^{96}$Zr can be determined based on the age of the measured zircon sample and the relative amount of the daughter product $^{96}$Mo compared to the parent $^{96}$Zr by applying Equations \ref{eq:ndn0} and \ref{eq:halflife}. The zircon sample was detrital, having a wide range of ages with a mean of 910(30) Ma based on N=510 measurements. The amount of Mo and Zr in the digested samples cannot be measured simultaneously by ICPMS as the smallest measurable Mo/Zr ratio was found to be 1:$10^6$. It was therefore required to measure the amounts separately, before and after the first ion exchange. This increased the uncertainty in this ratio to an estimated 20\% due to potential Mo loss during the ion exchange.

As the $^{96}$Mo excess for the zircons was found to be within 1$\sigma$ uncertainty of zero, only a lower limit can be placed on the half-life. As shown in Table \ref{tab:Halflife-Table}, despite the better precision of the 5655F-2 measurement, the larger amount of Mo lowers the limit that can be derived. The determined lower limit for the half-life of $^{96}$Zr is $T_{1/2} \geq 6.4\cdot10^{18}$.

\begin{table}[t]
	\centering
		\begin{tabular}{llll}
		\hline
		Sample & Mo/Zr & $\delta^{96}\text{Mo}$ & Half-life \\
		 & (ppm) & Upper limit & Lower limit \\
		\hline
		WSL-5655 (1) & 0.19 & 90 & $6.4 \cdot 10^{18}$ a \\
		WSL-5655 (2) & 2.4 & 42 & $1.1 \cdot 10^{17}$ a \\
		\hline
		\end{tabular}
	\caption{Lower limit of $^{96}$Zr half-life based on uncertainty of $\delta(^{96}\text{Mo})$.}
	\label{tab:Halflife-Table}
\end{table}

To directly measure half-life of $^{96}$Zr, a large high purity sample (a few grams) of older zircons with an age over 2 Ga would be required. The results of this study demonstrate that it will be possible to determine the half-life with similar precision to the previous direct counting-rates measurement\cite{Argyriades2010}. This procedure eliminates the need for assumptions used to estimate the non-radiogenic Mo content in zircon for the geochemical measurement by Wieser and DeLaeter\cite{Wieser2001}, which will allow an understanding of the discrepancy between these two measurements.

\section{Conclusions}

The techniques developed have demonstrated high precision measurements of trace Mo in ancient zircons. The chemical separation and data analysis techniques have allowed for high precision measurements to be made on very limited samples ($<50$ ng Mo from 0.5 g zircon) by reducing Zr content by $>9$ orders of magnitude while retaining 63\% of Mo. The flexibility of TEVA and cation resins, which combined have selectivity for most of the periodic table, allows these techniques to be applied to the measurement of other trace elements in zircon free of matrix effects and isobaric interferences. Further, the calibration of the ion exchange elution "online" allows for diagnostics and refinement of ion exchange procedures, optimizing acid volumes required for separation to maximize sample recovery while minimizing chemistry blank. The application of data filters and enhanced background and mass bias corrections have led to a significant improvement in measurement precision, often more than an order of magnitude improvement over typical data evaluation outputs. These algorithms can be applied to any isotopic system, particularly when isobaric interferences are present and sample is limited. Isotopic composition measurements of highly trace elements in ancient zircon will extend the application of state-of-the-art geochemical tracers to understanding early Earth evolution, long-lived nuclear processes and cosmochemistry.

\begin{acknowledgments}
We acknowledge the support of the Natural Sciences and Engineering Research Council of Canada (NSERC).
\end{acknowledgments}

\bibliography{Mendeley}

\end{document}